\documentclass[
reprint,
superscriptaddress,
 amsmath,amssymb,
 aps,
pra,
]{revtex4-1}

\usepackage{graphicx}
\usepackage{dcolumn}
\usepackage{bm}
\usepackage{caption}
\usepackage{subcaption}
\usepackage{physics}

\begin{document}

\title{Enabling CV-MDI-QKD for low squeezed states using non-Gaussian operations}

\author{Farsad Ahmad}
 \email{farsad.ahmad96@gmail.com}
 \affiliation{
 School of Natural Sciences, National University of Sciences and Technology, Islamabad, Pakistan
}
\author{Aeysha Khalique}
 \email{aeysha.khalique@sns.nust.edu.pk}
  \affiliation{
 School of Natural Sciences, National University of Sciences and Technology, Islamabad, Pakistan
}
  \affiliation{
National Centre for Physics (NCP), Shahdra Valley Road, Islamabad 44000, Pakistan.
}

\date{\today}

\begin{abstract}
We propose a new non-Gaussian version of continuous variables measurement device independent quantum key distribution (CV-MDI-QKD) protocol by utilizing photon added-then-subtracted (PAS) state. We report that our single and two mode PAS-CV-MDI-QKD protocols outperform pure state as well as two mode photon replaced (2PR) state CV-MDI-QKD protocols in the low squeezing and high noise regime, which is the practical regime. With such resources, at metropolitan distance of around 20km, CV-MDI-QKD is inaccessible with pure state and 2PR state, while PAS-CV-MDI-QKD can generate a useful key rate in this regime. Additionally we show that states with higher logarithmic negativity are not necessarily the best choice when used in CV-MDI-QKD.
\end{abstract}
\maketitle

\section{\label{sec:level1}INTRODUCTION}
Classical public key cryptography fails under a quantum computer attack \cite{shor}. Quantum key distribution (QKD) \cite{qkdreview1,qkdreview2} has emerged as a viable alternative, it promises unconditionally secure public key generation. The first proposed discrete variables (DV) QKD protocol was BB84 \cite{bb84} which has been experimentally demonstrated in several different environments \cite{airbb84}. BB84 is a discrete variables protocol that performs exceptional at long distances but is challenging to implement with current telecom infrastructure because it uses single photon resources. A more recent trend is focused on using continuous variables (CV) quantum signals \cite{cv2,cv3,cv1} which offer higher secure key rate while using existing telecom technology.

CV-QKD has been proven secure against arbitrary collective attacks in the asymptotic regime \cite{atk1,atk2}. CV-QKD schemes are completely secure as long as devices are ideal. However when device imperfections are considered, Eve, the adversary, can exploit security loopholes to attack instruments while staying hidden rendering the protocol insecure \cite{tap1,tap2,tap3,tap4,tap5}. Measurement device independent (MDI) \cite{dvmdi} and device independent \cite{dvdi} protocols resist these side-channel attacks where MDI offers a much more practical solution. Naturally, improving the performance of this protocol has become an exciting topic in CV-QKD.

Single photon non-Gaussian operations on two mode squeezed vacuum (TMSV) states such as a photon added-then-subtracted (PAS) operation and coherent superposition of photon addition and subtraction, which generates a photon replaced (PR) state, were first studied in \cite{pas} and \cite{rep1} respectively. Such non-Gaussian states have extensively been studied and produced in labs \cite{ngexp1,ngexp2,ngexp3,pasexp}. These states increase entanglement of Gaussian type states \cite{bartley} which can translate to better transmission distance and sometimes sustain more thermal noise when implemented in direct transmission scheme and in MDI setting \cite{zubairi1,mdisps,mdizpc,jsingh,zpc8mdi}.

Previously proposed non-Gaussian states when used in an MDI protocol require high squeezing of nearly $20$dB \cite{mdisps,mdizpc} to outperform the original CV-MDI protocol \cite{cvmdi}. A squeezing of approximately $15$dB is achievable in a close to ideal environment \cite{sqdb1,sqdb2} while a squeezing of $5$dB can easily be produced in labs \cite{sqdb3}. A practical MDI-QKD setting would have low squeezing and high noise, commercial realization of this protocol is not possible using pure states or PR states because when available squeezing is low, these states fail because of high noise. Our proposed protocol can overcome this boundary and enable positive key rate generation in this zone.

In this paper, we propose a new protocol based on PAS state which are generated by first adding a single photon and then removing it from the CV signal. We start by deriving a transmissivity dependant relation of PAS state and show that for a CV-MDI protocol using Gaussian TMSV states with initial squeezing less than $3$dB, these states lead to extended transmission distance in thermal loss channels and offer effective resilience to excess noise which pure state and PR state protocols fail to do. Compared to previously proposed zero photon catalysis (ZPC) MDI \cite{mdizpc} and single photon subtracted (SPS) MDI \cite{mdisps}, neither of these protocols outperform TMSV in this low squeezing regime.

PAS states that we have utilized in this paper are different from PR states, considered before in the same setup, in the sense that PAS states are generated by performing photon addition and subtraction sequentially. We also report that for the known PR state, higher logarithmic negativity as compared to PAS states, does not translate to better performance in terms of achievable distance and noise resistance.

The paper is structured as follows. In Sec~\ref{SEC2}, we describe the Gaussian and non-Gaussian CV-MDI-QKD protocol and derive their secret key rate under Gaussian collective attacks. In Sec.~\ref{SEC3}, we derive the transmissivity dependant relation of PAS states and show the performance of this state in MDI-QKD. Finally, the paper is concluded in Sec.~\ref{SEC4}.
\begin{figure*}
\begin{subfigure}{.5\textwidth}
  \includegraphics[width=1.4\linewidth]{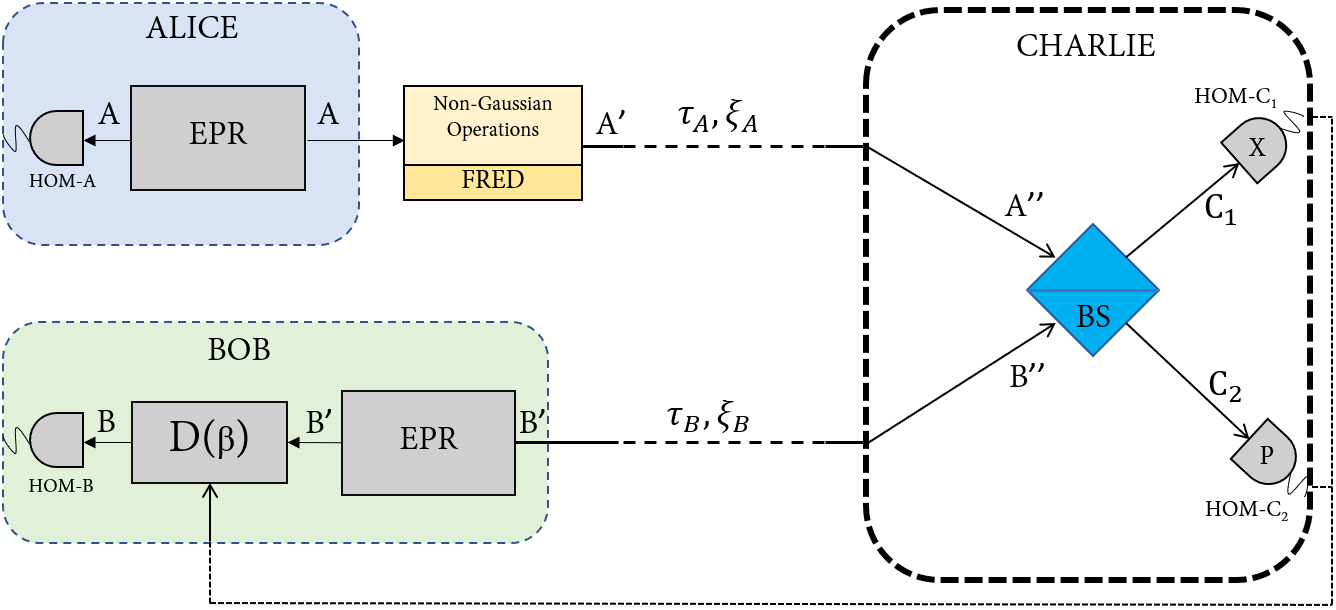}
  \caption{}
\end{subfigure}%
\begin{subfigure}{.5\textwidth}
\hspace{4cm}
  \subcaptionbox{}{
\vspace{0.4in}\includegraphics[width=0.5\linewidth]{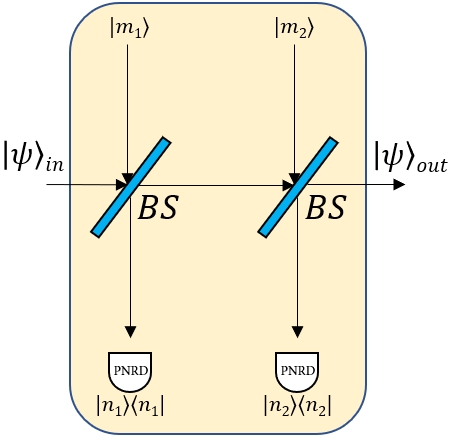}}
\end{subfigure}
\caption{\label{fig1} (a) Entanglement based non-Gaussian CV-MDI-QKD scheme. $D(\beta)$ is the displacement operator and $\tau_A$($\tau_B$) and $\xi_A$($\xi_B$) are channel transmissivities and excess noises for Alice and Bob respectively. (b) Non-Gaussian operation at Fred's station. $\ket{\psi}_{\text{1PAS}}$ is created by setting $m_1 = n_2 = 1, m_2 = n_1 = 0$ when $\ket{\psi}_{\text{in}} =\ket{\psi}_{\text{TMSV}}$, $\ket{\psi}_{\text{2PAS}}$ is created by setting $m_1 = n_2 = 1, m_2 = n_1 = 0$ when $\ket{\psi}_{\text{in}} =\ket{\psi}_{\text{1PAS}}$ and $\ket{\psi}_{\text{2PR}}$ is created by setting $m_1 = n_1 = m_2 = n_2 = 1$ when $\ket{\psi}_{\text{in}} =\ket{\psi}_{\text{TMSV}}$.}
\end{figure*}
\begin{figure}
  \fbox{\includegraphics[width=0.95\linewidth]{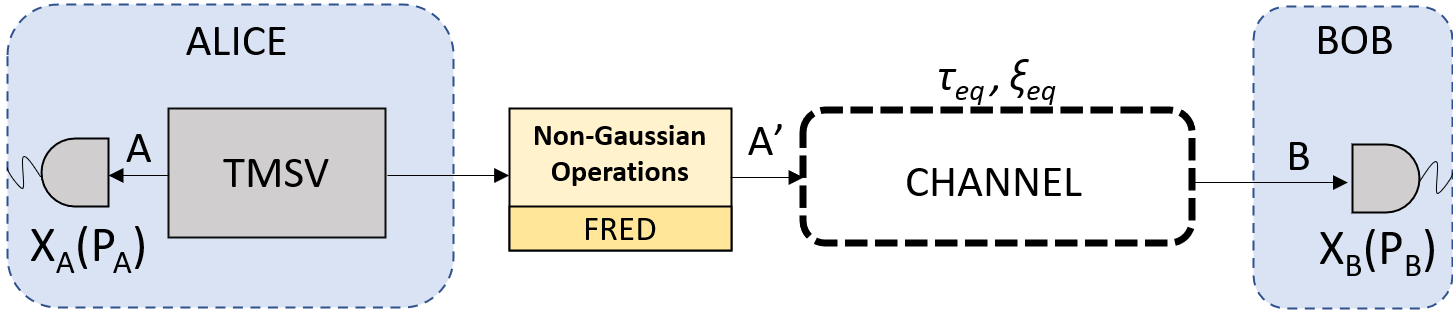}}
  \caption{\label{eqscheme}Equivalent one-way CV-MDI-QKD protocol in the entanglement scheme with specified non-Gaussian operations performed by Fred. The setup is under the assumption that Eve controls Charlie, Bob's EPR state and the displacement except the homodyne detectors at Bob's station. $\tau_{eq}$ and $\xi_{eq}$ are the equivalent thermal loss and excess noise of the channel}
\end{figure}
\section{GAUSSIAN AND NON-GAUSSIAN CV-MDI-QKD PROTOCOL\label{SEC2}}
In this section we provide the necessary background for Gaussian and non-Gaussian CV-MDI-QKD protocols. First we present a short summary of the original entanglement based (EB) CV-MDI-QKD protocol \cite{cvmdi} in Sec.~\ref{SEC2A}. We then show the non-Gaussian version of this protocol with PR operations in Sec.~\ref{SEC2B}. We then show the reduction of this protocol to a simpler prepare and measure setup, we express the secret key rate for this protocol under Gaussian collective attacks and discuss logarithmic negativity in Sec.~\ref{SEC2C}.
\subsection{Entanglement based Gaussian CV-MDI-QKD\label{SEC2A}}

Fig. \ref{fig1}(a), with the omission of Fred, represents EB Gaussian CV-MDI-QKD protocol with homodyne detection. In the EB scheme of CV-MDI-QKD protocol, Alice generates a two-mode squeezed vacuum state (TMSV) with initial variance $V_A$. She keeps the mode $A$ and sends the other mode to an untrusted party named Charlie through a thermal loss channel with transmissivity $\tau_A$ and excess noise $\xi_A$, where excess noise is the sum of all forms of losses apart from the channel loss. Similarly, Bob also generates a TMSV with initial variance $V_B$, keeps the mode $B'$ and sends the other one to Charlie through his channel of transmissivity $\tau_B$ and excess noise $\xi_B$. Charlie receives the lossy states $A''$ and $B''$ and implements optimal Gaussian entanglement swapping by interfering the two incoming modes on a balanced beam splitter (BS) followed by a position and momentum homodyne measurement on the outgoing modes $C_1$ and $C_2$. He then announces the results of measurement through a classical channel $\{X_{C_1},P_{C_2}\}$. Bob then applies a displacement operator $D(\beta)$ with $\beta = gX_{C_1} + igP_{C_2}$ on his mode $B'$ which translates it to $B$, where $g$ is a classical gain factor. This creates correlation between Bob's mode $B$ and Alice's mode $A$. Both parties perform a homodyne measurement on their respective modes and continue to post processing.

Alice and Bob start by creating an entangled state which has the following covariance matrix
\begin{align}   
\Gamma_{A} &=
\begin{pmatrix}
a_1 \mathbf{I} & c \ \textbf{Z} \\
c \ \textbf{Z}& a_1 \mathbf{I}
\end{pmatrix} \ \ \textbf{;} \ \ 
\Gamma_{B'} =
\begin{pmatrix}
b_1 \mathbf{I} & d \ \textbf{Z} \\
d \ \textbf{Z}& b_1 \mathbf{I}
\end{pmatrix},\label{eqn:generalformstart}
\end{align}
where \textbf{I} = diag[1,1], \textbf{Z} = diag[1,-1]. Here, $a_1$ is the preparation variance of Alice's state, $c$ is the covariance of Alice's state, $b_1$ is the preparation variance of Bob's state and $d$ is the covariance of Bob's state.

In the Gaussian version of this protocol, Alice and Bob use a two mode squeezed vacuum (TMSV) state which gives $a_1 = V_A$, $c = \sqrt{V_A^2 - 1}$, $b_1 = V_B$ and $d = \sqrt{V_B^2 - 1}$.

\begin{figure*}
\begin{subfigure}{.5\textwidth}
  \subcaptionbox{}{
    \includegraphics[width=1\linewidth]{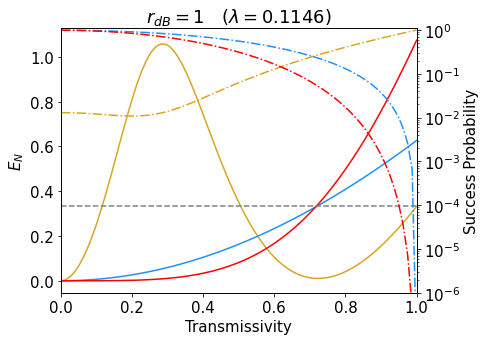}}
\end{subfigure}%
\begin{subfigure}{.5\textwidth}
  \subcaptionbox{}{
    \includegraphics[width=1\linewidth]{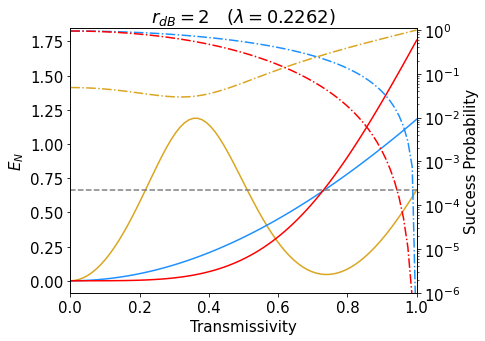}}
\end{subfigure}
  \caption{\label{fig-logN}(Color online) Logarithmic negativity $E_\text{N}$(solid line) and success probability (dashed dotted line) of 1PAS state (blue), 2PAS state (red) and 2PR state (brown) as a function of beam splitter transmissivity $T$ at initial squeezing of (a) $r_{\text{dB}} = 1$dB and (b) $r_{\text{dB}} = 2$dB. $E_\text{N}$ of pure TMSV is graphed as the gray dashed line.}
\end{figure*}

For the extreme asymmetric setting $(\tau_B = 1)$, implementing channel loss with excess noise independently on the traveling modes of Alice ($A''$) and Bob ($B''$) results in the following covariance matrices of Alice ($\Gamma_{A''}$) and Bob ($\Gamma_{B''}$)
\begin{align}   
\Gamma_{A''} &=
\begin{pmatrix}
a_1 \mathbf{I} & \sqrt{\tau_A}c \ \textbf{Z} \\
\sqrt{\tau_A}c \ \textbf{Z}& a_2 \mathbf{I}
\end{pmatrix} \ \ \textbf{;} \ \ 
\Gamma_{B''} =
\begin{pmatrix}
b_2 \mathbf{I} & d \ \textbf{Z} \\
d \ \textbf{Z}& b_1 \mathbf{I}
\end{pmatrix},\label{eq:losses}
\end{align}
where $a_2$ and $b_2$ are variances of the lossy modes received by Charlie given as,
\begin{align}
a_2 &= \tau_A (a_1 -1) + 1 + \xi_A ,\\ 
b_2 &= b_1 + \xi_B.
\end{align}
Here $\xi_A$($\xi_B$) is the excess noise in Alice (Bob) channel and $\tau_A$ is Alice-Charlie channel transmittance which is related to the distance between them ($L$) as $\tau_A = 10^{(-\alpha L/10)}$. $\alpha$ is the fixed loss attenuation factor typically set at 0.2dB/km. In our simulations we assume $\xi_A = \xi_B$ and we define total excess noise as $\xi := \xi_A + \xi_B$.

An optimal Gaussian entanglement swapping between the second mode of $\Gamma_{A''}$ and the first mode of $\Gamma_{B''}$, as performed by Charlie, will result in the following covariance matrix \cite{phdthesis}
\begin{align}
\label{eqn:mdi}
\Gamma_{AB} &=
\begin{pmatrix}
x_1 \mathbf{I} & x' \ \textbf{Z} \\
x' \ \textbf{Z}& x_2 \mathbf{I}
\end{pmatrix},
\end{align}
where variance of modes sitting in Alice's (Bob's) lab is now $x_1$ ($x_2$) and their covariance is $x'$ given as
\begin{align}
\label{eqn6}x_1 &= a_1 - (\tau_Ac^2)/(a_2+b_2)  ,\\
\label{eqn7}x_2 &= b_1 - d^2/(a_2+b_2)  ,\\
\label{eqn8}x' &= (cd\sqrt{\tau_A})/(a_2 + b_2) .
\end{align}
This forms the final covariance matrix of states shared by Alice and Bob which we will use for the extraction of key rate.

\subsection{Photon replaced non-Gaussian CV-MDI-QKD\label{SEC2B}}

Now we consider a non-Gaussian version of this protocol. We assume that Alice is the sender and Bob is the receiver of information and their reference of reconciliation is Bob. We also assume that Charlie is very close to Bob, such that $\tau_B = 1$. These assumptions imply that higher entanglement at Alice's station will result in better performance, so we apply our non-Gaussian operations on Alice's outgoing state.

In our MDI setting as seen in Fig. \ref{fig1}(a), Fred is responsible to perform the chosen single photon non-Gaussian operation on Alice's TMSV state. This operation is performed by mixing the incoming state $\ket{\psi}_{\text{in}}$ with a single photon state $\ket{m_1}$ on a beam splitter (BS) of transmissivity $T$. One part of the outgoing state is projected on a single photon detector $\ket{n_1}\bra{n_1}$ and the other half of the state goes through another BS with the same transmissivity $T$ mixing it with a single photon state $\ket{m_2}$. One half of this state is projected on $\ket{n_2}\bra{n_2}$ while the other half is our outgoing state $\ket{\psi}_{\text{out}}$. This sequence of operations generate the following outgoing state
\begin{align}\label{eq:2}
\ket{\psi}^{\text{out}} = \hat{\Pi}_{n_2,d}\hat{U}_{bd}(T)\hat{\Pi}_{n_1,c}\hat{U}_{bc}(T) \big(\ket{\psi}_{ab}^{\text{in}}\otimes \ket{m_1m_2}_{cd}\big),
\end{align}
where $\hat{\Pi}_{n_p,i} = \big[\ket{n_p}\bra{n_p}\big]_i$ is the photon number resolving operation, $\hat{U}_{ij}(T) = \text{exp}\big[\text{arccos}(\sqrt{T})(\hat{a}^\dag_i\hat{a}_j - \hat{a}_i\hat{a}^\dag_j)\big]$ is the beam splitter operator with $p \in \{1,2\}$ and $i,j \in \{a,b,c,d\}$ corresponding to the mode on which the operators act. We represent $\{\ket{m_p},\ket{n_p}\}_{ij}$ in the Fock basis for modes $i$ and $j$.

We can generate a two-mode photon replaced (2PR) state by setting $m_1 = n_1 = 1$ which mixes 1 photon with the incoming beam and detects 1 photon from one part of the state. Setting $m_2 = n_2 = 1$ does the same operation again. We take $\ket{\psi}_{\text{in}} = \ket{\psi}_{\text{TMSV}}$ where TMSV is the two mode squeezed vacuum state. TMSV state is created in the photon number space by squeezing two independent vacua $\ket{00}_{ab}$. The TMSV state has the following form
\begin{equation}\ket{\psi}_{\text{TMSV}} = \hat{S}(r)\ket{00}_{ab}= \sum_{n=0}^\infty\sqrt{1-\lambda^2} \lambda^n \ket{nn}_{ab}.\end{equation}
$\hat{S}(r)$ is the two mode squeezing operator defined as $\hat{S}(r) = \text{exp}\big(r(\hat{a}_a\hat{a}_b - \hat{a}_a^\dag \hat{a}_b^\dag)\big)$, $\lambda = \text{tanh}(r)$ and $r$ is the squeezing parameter which we convert to decibel using $r_{\text{dB}} = 10\ \text{log}_{10}(e^{2r})$.
The 2PR state $\ket{\psi}_{\text{2PR}}$ thus generated is
\begin{equation}\ket{\psi}_{\text{2PR}} = \mathcal{N}_{\text{2PR}}\sum_{n=0}^\infty\sqrt{1-\lambda^2} \lambda^n T^{2n-2} \big[T^2-n(1-T^2)\big]^2 \ket{nn},\end{equation}
where $\mathcal{N}_{\text{2PR}}$ is the normalization constant and $\mathcal{P}_{\text{2PR}}$ is the probability of success of two photon replacement operation, which depends on the squeezing $(\lambda)$ of the initial state and transmissivity $(T)$ of the mixing beam splitters as
\begin{align}
\frac{1}{(\mathcal{N}_{\text{2PR}})^2} = \mathcal{P}_{2PR} &= \frac{1-\lambda^2}{(1-\lambda^2T^4)^5}\Big[ \lambda^{6} T^{8} (T^{8} - 8 T^{6} + \nonumber\\& 24 T^{4} - 32 T^{2} + 11) + T^{4} \lambda^{4} (11 T^{8} -\nonumber\\& 56 T^{6} +  96 T^{4} - 56 T^{2} + 11) + T^{4} + \nonumber\\& \lambda^{2} (11 T^{8} - 32 T^{6} + 24 T^{4} - 8 T^{2} + \nonumber\\&1)^{8} + T^4 + \lambda T^{12}
\Big].
\end{align}
For a given state $\ket{\psi}$, we can get Alice's covariance matrix by using 
\begin{align}
\label{eqn:cmforng}
\Gamma_A = V_{ij} = \langle \{ \Delta \hat{x}_i , \Delta \hat{x}_j \} \rangle
\end{align}
where \{.\} is the anti-commutator, $\hat{x}_i \in \{\hat{q}_i,\hat{p}_i\}$, $\Delta \hat{x}_i = \hat{x}_i - \langle \hat{x}_i \rangle$ and $\langle . \rangle = \bra{\psi} . \ket{\psi}$. This will generate a non-Gaussian covariance matrix in the from of Eq.~\eqref{eqn:generalformstart}, from there we follow Eq.~\eqref{eq:losses} to implement channel losses followed by using the relation given in Eq.~\eqref{eqn:mdi} to implement entanglement swapping which then gives us the final covariance matrix shared by Alice and Bob.
\begin{figure*}
\begin{subfigure}{.47\textwidth}
\hspace{-1.0cm}
  \subcaptionbox{}{
    \includegraphics[width=1\linewidth]{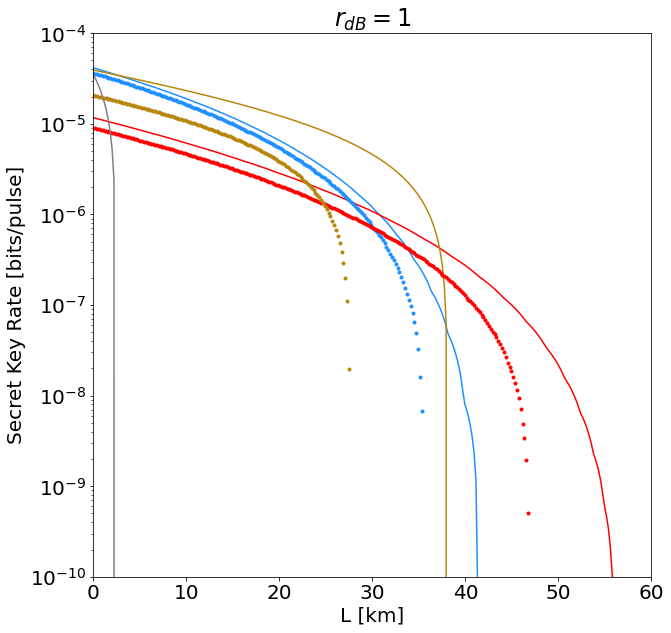}}
\end{subfigure}
\begin{subfigure}{.47\textwidth}
  \subcaptionbox{}{
    \includegraphics[width=1\linewidth]{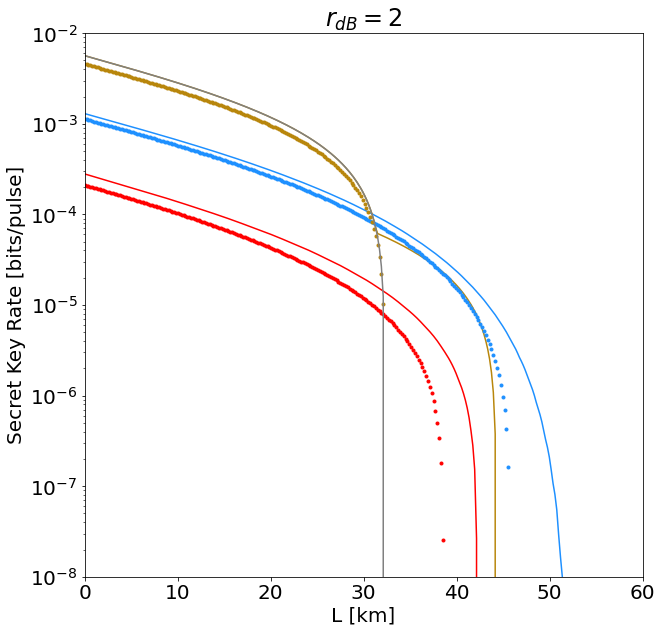}}
\end{subfigure}
  \caption{\label{fig:keyatrdb}(Color online) SKR of extreme asymmetric CV-MDI-QKD as a function of Alice-Charlie distance with fixed $\xi = 0.004$, reconciliation efficiency $\gamma = 95$\% and optimized over beam splitter transmissivity $T$ at two different values of initial squeezing. Ideal conditions are plotted as solid lines and practical practical conditions are plotted as dotted lines, with PNRD efficiency set to $95$\%. (a)~$r_{\text{dB}} = 1$dB and (b)~$r_{\text{dB}} = 2$dB. TMSV (Gray), 1PAS state (Blue), 2PAS state (Red) and 2PR state (Brown).}
\end{figure*}
\subsection{Security proof, key rate and log-negativity\label{SEC2C}}

The CV-MDI protocol has two quantum channels, one transmits Alice's state to Charlie and the other transmits Bob's state to Charlie. There are two ways Eve can attack these channels, one mode attack through an entangling cloner attack on each channel independently, and a two mode attack through a correlated Gaussian coherent attack on both channels. While a two mode attack is the more general attack model, in a practical setup we can restrict our security to a one mode Gaussian collective attack because we are taking an asymmetric setting where Bob-Charlie distance is practically zero.

When we offer Eve more control, we assume that Charlie is entirely under Eve's influence, in addition to that, Bob's displacement operation as well as Bob's EPR state is under Eve's influence as well. In this setting, only Bob's homodyne detector is trusted in his lab. The setup is shown in Fig.~\ref{eqscheme}, these assumptions reduce PAS-CV-MDI-QKD to a simpler one-way CV-QKD with PAS states
in the entanglement scheme. This new channel has equivalent parameters $\tau_{\text{eq}}$ and $\xi_{\text{eq}}$. To minimize $\xi_{\text{eq}}$, $g$ is taken as $g^2 = \frac{2V_B - 2}{\tau_B(V_B + 1)}$, then the equivalent channel parameters become \cite{cvmdi}
\begin{align}
\tau_{\text{eq}} &= \frac{1}{2}\tau_Ag^2, \\
\xi_{\text{eq}} &= \xi_A + \frac{\tau_B}{\tau_A}\big(\xi_B-2 \big) + \frac{2}{\tau_A}.
\end{align}

The number of secure bits generated from each pulse is called the secret key rate (SKR). The key rate with non-Gaussian operations can be evaluated using
\begin{equation}
    \text{SKR} = \mathcal{P}(\gamma I_{\text{AB}} - \chi_{\text{BE}}), \label{eqn:skr}
\end{equation}
where $\gamma$ is the efficiency of reverse reconciliation, $I_{\text{AB}}$ is Alice and Bob's mutual Shannon information, $\chi_{\text{BE}}$ is the Eve-Bob Holevo bound and $\mathcal{P}$ is the probability of success of the non-Gaussian operation that Fred applies. From this relation, only those pulses for which the non-Gaussian operation is successful are kept and rest are discarded in post processing.
From the final covariance matrix in standard form Eq.~\eqref{eqn:mdi}, $I_{\text{AB}}$ with homodyne measurement at both stations is
\begin{equation}
    I_{\text{AB}} = \frac{1}{2}\text{log}_2\Big(\frac{x_2}{x_2 - (x'^2/x_1)}\Big).
\end{equation}
where $x_1$, $x_2$ and $x'$ are as given in Eq.~(\ref{eqn6}$-$\ref{eqn8}). Since Eve holds a purification of Alice-Bob state, we can evaluate Eve's upper bound from the symplectic eigen values $v_1$ and $v_2$ of $\Gamma_{\text{AB}}$ and the symplectic eigen value $\bar{v}$ of Eve's conditional covariance matrix when Bob performs homodyne measurement. These eigen values are 
\begin{align}
v_{1,2} &= \sqrt{\frac{\Delta}{2} \pm \frac{\sqrt{\Delta^2 - 4 \  \abs{\Gamma_{AB}}}}{2}},
\\
\bar{v} &= \sqrt{x_1^2-\frac{x_1x'^2}{x_2}}.
\end{align}
$\Delta = x_1^2 + x_2^2 - 2x'^2$ and $\abs{\Gamma_{AB}}$ is the determinant of covariance matrix shared by Alice and Bob. We can then get the upper bound on Eve's information using $\chi_{\text{BE}} = G(v_1) + G(v_2) - G(\bar{v})$, where $G(x) = \frac{1+x}{2}\ \text{log}_2\frac{1+x}{2} - \frac{x-1}{2}\ \text{log}_2\frac{x-1}{2}$.
\begin{figure*}
\begin{subfigure}{.47\textwidth}
\hspace{-1.0cm}
  \subcaptionbox{}{
    \includegraphics[width=1\linewidth]{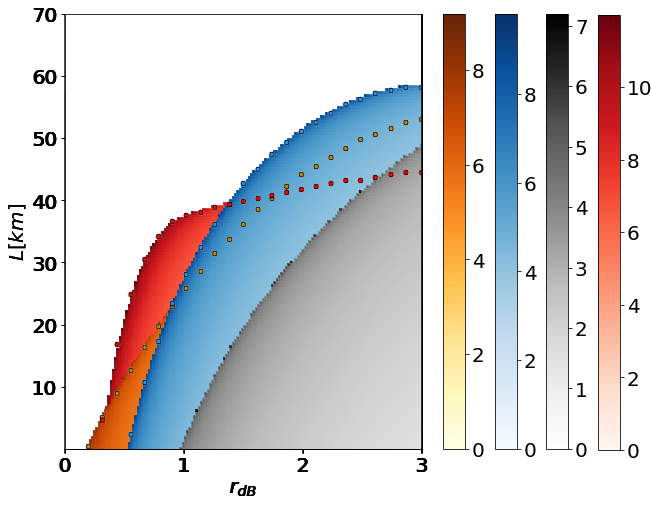}}
\end{subfigure}
\begin{subfigure}{.47\textwidth}
  \subcaptionbox{}{
    \includegraphics[width=1\linewidth]{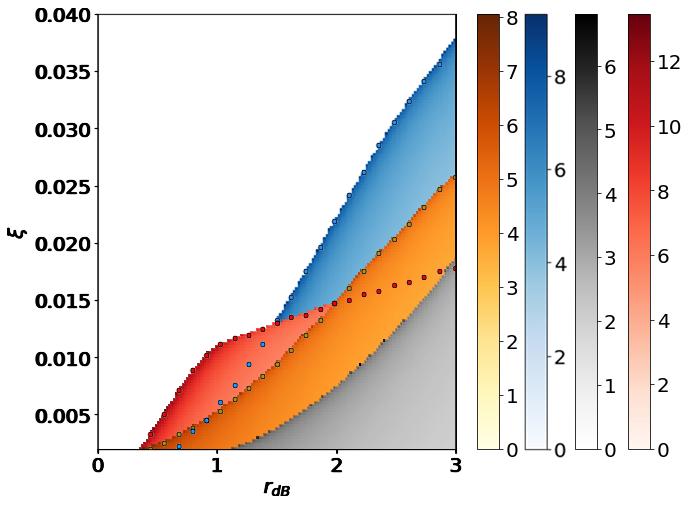}}
\end{subfigure}
  \caption{\label{fig:heatmap}(Color online) Heat map of -log$_{10}$(SKR) of CV-MDI-QKD protocol set at $\gamma = 0.95$ as a function of initial squeezing $r_{\text{dB}}$ and (a) channel length from Alice to Charlie with fixed $\xi = 0.004$, (b) total excess noise $\xi$ with fixed channel length at 20km. TMSV (Gray), 1PAS state (Blue), 2PAS state (Red) and 2PR state (Brown).}
\end{figure*}
Next we look at the entanglement log-negativity \cite{logn1,logn2} of these non-Gaussian states. Logarithmic negativity is an entanglement monotone defined as 
\begin{equation}
E_\text{N} = \text{log}_2\big( \norm{\rho^{\text{PT}}}\big)    
\end{equation}
where $\rho^{\text{PT}}$ denotes the partial transpose of our mixed state $\rho = \ket{\psi}\bra{\psi}$, $\ket{\psi}$ is the outgoing state from Fred's station and $\norm{.}$ is the trace norm operation.

This quantity puts an upper bound on the amount of entanglement that can be distilled from a mixed state with higher values of $E_\text{N}$ corresponding to larger entanglement distillation capacity.

\section{PAS-CV-MDI-QKD\label{SEC3}}
In this section we show the utility of photon added-then-subtracted (PAS) state as the primary entanglement resource in MDI setting. We first derive a transmissivity dependent PAS state in Sec.~\ref{SEC3A}. Next we show the performance of this state compared with 2PR state and pure TMSV in MDI-QKD in Sec~\ref{SEC3B}.
\subsection{PAS states\label{SEC3A}}
A PAS state is generated by first applying the photon addition operation on the incoming beam followed by a photon subtraction operation. We can apply this operation if we set $m_1 = n_2 = 1$ and $m_2 = n_1 = 0$ in Eq.~\eqref{eq:2}. This sequence first adds a photon in the incoming beam and then subtracts a photon from it.

We can set $\ket{\psi}_{\text{in}} = \ket{\psi}_{\text{TMSV}}$ which produces 1PAS state $\ket{\psi}_{\text{1PAS}}$ or we can set $\ket{\psi}_{\text{in}} = \ket{\psi}_{\text{1PAS}}$ which produces 2PAS state $\ket{\psi}_{\text{2PAS}}$. The states generated are
\begin{align}\ket{\psi}_{\text{1PAS}} &= \mathcal{N}_{1}\sum_{n=0}^\infty\sqrt{1-\lambda^2} \lambda^n T^{2n} (1-T^2)(n+1) \ket{nn} ,\\
\ket{\psi}_{\text{2PAS}} &= \mathcal{N}_{2}\sum_{n=0}^\infty\sqrt{1-\lambda^2} \lambda^n T^{4n} (1-T^2)^2(n+1)^2 \ket{nn},
\end{align}
where $\mathcal{N}_1$ and $\mathcal{N}_2$ are the normalization constants which are related to success probabilities $\mathcal{P}_1$ and $\mathcal{P}_2$ as
\begin{align}
\frac{1}{(\mathcal{N}_{1})^2} = \mathcal{P}_{1} &= \frac{(1-\lambda^2)(1-T^2)^2 \big(\zeta_1 + 1 \big)}{(1-\zeta_1)^3},\\
\frac{1}{(\mathcal{N}_{2})^2} = \mathcal{P}_{2} &= (1-\lambda^2)(1-T^2)^4 \times \nonumber \\&\Bigg[\frac{-16\zeta_2^4 -\zeta_2^3 -11\zeta_2^2 +5\zeta_2 - 1}{\zeta_2^5 -5\zeta_2^4 + 10\zeta_2^3 -10\zeta_2^2 + 5\zeta_2 - 1}\Bigg],
\end{align}
where $\zeta_1 =\lambda^2T^4$ and $\zeta_2 = \lambda^2T^8$. 

For these two states we can get Alice's covariance matrix $\Gamma_A$ using Eq.~\eqref{eqn:cmforng}. Once we have $\Gamma_A$ we use Eq.~\eqref{eq:losses} to implement channel and thermal losses and finally we apply an entanglement swapping using Eq.~\eqref{eqn:mdi}. This final covariance matrix is then used for the extraction of secret key rate. Note that only Alice's state is non-Gaussian whereas Bob is using a pure TMSV state.

In our simulations, for each realization of Alice-Charlie distance $L$ and excess noise $\xi$ we optimize the covariance matrix over beam splitter transmissivity $T$ which maximizes the distance $L$ and noise resilience $\xi$.

\subsection{Performance\label{SEC3B}}

Fig.~\ref{fig-logN}(a) and Fig.~\ref{fig-logN}(b) show the amount of distillable entanglement (solid line) and probability of success (dashed dotted line) for 2PR (brown), 1PAS (blue) and 2PAS (red) states alongside pure TMSV state (gray) as a function of beam splitter transmissivity $T$. We have plotted log-negativity $E_\text{N}$ and success probabilities $\mathcal{P}_1$, $\mathcal{P}_2$ and $\mathcal{P}_{\text{2PR}}$ for two different values of initial squeezing i.e. $r_{\text{dB}} = 1$dB and $r_{\text{dB}} = 2$dB.

Fig.~\ref{fig:keyatrdb}(a) and Fig.~\ref{fig:keyatrdb}(b) show the plot of the secret key rate when excess noise is set to $\xi=0.004$ for 2PR (brown), 1PAS (blue) and 2PAS (red) states alongside pure TMSV state (gray) as a function of Alice-Charlie distance $L$. We have plotted SKR for two different values of initial squeezing $r_{\text{dB}} = 1$dB and $r_{\text{dB}} = 2$dB.

From Fig.~\ref{fig-logN}(a) we can observe that the maximum $E_\text{N}$ of 2PR state is almost equal to 2PAS state at $r_{\text{dB}}=1$dB, while our results for CV-MDI-QKD given in Fig.~\ref{fig:keyatrdb}(a) suggest that 2PAS state can achieve a longer transmission distance than 2PR state. Similarly in Fig.~\ref{fig-logN}(b) we observe that the maximum $E_\text{N}$ of 2PAS state is above 1PAS state at $r_{\text{dB}} = 2$dB while Fig.~\ref{fig:keyatrdb}(b) suggest that 1PAS state would outperform 2PAS and 2PR state in terms of transmission distance.

With these results, a direct correlation between higher values of $E_\text{N}$ and a better performance in MDI cannot be established for such non-Gaussian states. A previous work focused on low squeezed states for QKD \cite{lowsq} has also noted that their states with higher $E_\text{N}$ were outperformed by states with lower $E_\text{N}$.

Next we simulate the performance of all three non-Gaussian states alongside pure TMSV state in two different settings. Fig.~\ref{fig:heatmap} is a heat map of $-\text{log}_{10}(\text{SKR})$ plotted when initial squeezing of TMSV states is taken between $0$dB and $3$dB for two different cases: Fig.~\ref{fig:heatmap}(a) when total excess noise $\xi$ is fixed at 0.004 and channel length between Alice and Charlie is varied, Fig.~\ref{fig:heatmap}(b) when channel length between Alice and Charlie is fixed to 20km and excess noise is varied.

When we look at Fig.~\ref{fig:heatmap}(a) and Fig.~\ref{fig:heatmap}(b), we observe that below $r_{\text{dB}} \approx 1.5$dB, 2PAS-CV-MDI-QKD is the best choice for longer transmission distance and noise resilience while above $r_{\text{dB}} \approx 1.5$dB, 1PAS-CV-MDI-QKD outperform all other states in terms of transmission distance and noise resilience. 1PAS-CV-MDI-QKD is a general improvement over 2PR state except that 2PR-CV-MDI-QKD offers a slightly higher key rate.

\section{CONCLUSION\label{SEC4}}
We have proposed a new non-Gaussian entanglement based photon added-then-subtracted continuous variables measurement device independent protocol, PAS-CV-MDI-QKD, under extreme asymmetric setting that can enhance transmission distance and offer flexibility in excess noise while working with low squeezed two mode squeezed vacuum (TMSV) state. These non-Gaussian states are produced by applying PAS and photon replacement (PR) operations on TMSV states. PAS operation is applied once on a pure TMSV state to produce 1PAS state and again on a 1PAS state to produce a 2PAS state. PR operation is applied once on a 1PR state to produce 2PR state. To the best of our knowledge, PAS states have not been used for QKD with continuous variables before.

These non-Gaussian states increase entanglement of a pure TMSV state, but we observe that a direct correlation between having a higher value of maximum distillable entanglement does not translate to a better performing MDI protocol.

To highlight the advantages offered by our PAS-CV-MDI-QKD protocol, for 1PAS state we observe that it can generate a positive key rate at metropolitan distances for low squeezing and high thermal noise. For 2PAS state we observe that it offers the longest transmission distance when initial squeezing of TMSV is less than $\approx 1.5$dB and a 2PR state was observed to have a small region of applicability, it offered slightly better key rate than 1PAS-CV-MDI-QKD and 2PAS-CV-MDI-QKD states but the general performance of 1PAS state was above it.

While previously proposed single photon subtracted (SPS) MDI \cite{mdisps} and zero photon catalysis (ZPC) MDI \cite{mdizpc} have the advantage of achieving longer transmission distance, both of these protocols fail to outperform pure state when squeezing is below 3dB. The two PAS states proposed here overcome this limitation.

\appendix
\nocite{*}
\bibliography{draft}

\begin{thebibliography}{38}%
\makeatletter
\providecommand \@ifxundefined [1]{%
 \@ifx{#1\undefined}
}%
\providecommand \@ifnum [1]{%
 \ifnum #1\expandafter \@firstoftwo
 \else \expandafter \@secondoftwo
 \fi
}%
\providecommand \@ifx [1]{%
 \ifx #1\expandafter \@firstoftwo
 \else \expandafter \@secondoftwo
 \fi
}%
\providecommand \natexlab [1]{#1}%
\providecommand \enquote  [1]{``#1''}%
\providecommand \bibnamefont  [1]{#1}%
\providecommand \bibfnamefont [1]{#1}%
\providecommand \citenamefont [1]{#1}%
\providecommand \href@noop [0]{\@secondoftwo}%
\providecommand \href [0]{\begingroup \@sanitize@url \@href}%
\providecommand \@href[1]{\@@startlink{#1}\@@href}%
\providecommand \@@href[1]{\endgroup#1\@@endlink}%
\providecommand \@sanitize@url [0]{\catcode `\\12\catcode `\$12\catcode
  `\&12\catcode `\#12\catcode `\^12\catcode `\_12\catcode `\%12\relax}%
\providecommand \@@startlink[1]{}%
\providecommand \@@endlink[0]{}%
\providecommand \url  [0]{\begingroup\@sanitize@url \@url }%
\providecommand \@url [1]{\endgroup\@href {#1}{\urlprefix }}%
\providecommand \urlprefix  [0]{URL }%
\providecommand \Eprint [0]{\href }%
\providecommand \doibase [0]{http://dx.doi.org/}%
\providecommand \selectlanguage [0]{\@gobble}%
\providecommand \bibinfo  [0]{\@secondoftwo}%
\providecommand \bibfield  [0]{\@secondoftwo}%
\providecommand \translation [1]{[#1]}%
\providecommand \BibitemOpen [0]{}%
\providecommand \bibitemStop [0]{}%
\providecommand \bibitemNoStop [0]{.\EOS\space}%
\providecommand \EOS [0]{\spacefactor3000\relax}%
\providecommand \BibitemShut  [1]{\csname bibitem#1\endcsname}%
\let\auto@bib@innerbib\@empty
\bibitem [{\citenamefont {Shor}(1994)}]{shor}%
  \BibitemOpen
  \bibfield  {author} {\bibinfo {author} {\bibfnamefont {P.~W.}\ \bibnamefont
  {Shor}},\ }in\ \href@noop {} {\emph {\bibinfo {booktitle} {Proceedings 35th
  annual symposium on foundations of computer science}}}\ (\bibinfo
  {organization} {Ieee},\ \bibinfo {year} {1994})\ pp.\ \bibinfo {pages}
  {124--134}\BibitemShut {NoStop}%
\bibitem [{\citenamefont {Scarani}\ \emph {et~al.}(2009)\citenamefont
  {Scarani}, \citenamefont {Bechmann-Pasquinucci}, \citenamefont {Cerf},
  \citenamefont {Du\ifmmode~\check{s}\else \v{s}\fi{}ek}, \citenamefont
  {L\"utkenhaus},\ and\ \citenamefont {Peev}}]{qkdreview1}%
  \BibitemOpen
  \bibfield  {author} {\bibinfo {author} {\bibfnamefont {V.}~\bibnamefont
  {Scarani}}, \bibinfo {author} {\bibfnamefont {H.}~\bibnamefont
  {Bechmann-Pasquinucci}}, \bibinfo {author} {\bibfnamefont {N.~J.}\
  \bibnamefont {Cerf}}, \bibinfo {author} {\bibfnamefont {M.}~\bibnamefont
  {Du\ifmmode~\check{s}\else \v{s}\fi{}ek}}, \bibinfo {author} {\bibfnamefont
  {N.}~\bibnamefont {L\"utkenhaus}}, \ and\ \bibinfo {author} {\bibfnamefont
  {M.}~\bibnamefont {Peev}},\ }\href {\doibase 10.1103/RevModPhys.81.1301}
  {\bibfield  {journal} {\bibinfo  {journal} {Rev. Mod. Phys.}\ }\textbf
  {\bibinfo {volume} {81}},\ \bibinfo {pages} {1301} (\bibinfo {year}
  {2009})}\BibitemShut {NoStop}%
\bibitem [{\citenamefont {Gisin}\ \emph {et~al.}(2002)\citenamefont {Gisin},
  \citenamefont {Ribordy}, \citenamefont {Tittel},\ and\ \citenamefont
  {Zbinden}}]{qkdreview2}%
  \BibitemOpen
  \bibfield  {author} {\bibinfo {author} {\bibfnamefont {N.}~\bibnamefont
  {Gisin}}, \bibinfo {author} {\bibfnamefont {G.}~\bibnamefont {Ribordy}},
  \bibinfo {author} {\bibfnamefont {W.}~\bibnamefont {Tittel}}, \ and\ \bibinfo
  {author} {\bibfnamefont {H.}~\bibnamefont {Zbinden}},\ }\href {\doibase
  10.1103/RevModPhys.74.145} {\bibfield  {journal} {\bibinfo  {journal} {Rev.
  Mod. Phys.}\ }\textbf {\bibinfo {volume} {74}},\ \bibinfo {pages} {145}
  (\bibinfo {year} {2002})}\BibitemShut {NoStop}%
\bibitem [{\citenamefont {Bennett}\ and\ \citenamefont
  {Brassard}(1984)}]{bb84}%
  \BibitemOpen
  \bibfield  {author} {\bibinfo {author} {\bibfnamefont {C.~H.}\ \bibnamefont
  {Bennett}}\ and\ \bibinfo {author} {\bibfnamefont {G.}~\bibnamefont
  {Brassard}},\ }in\ \href@noop {} {\emph {\bibinfo {booktitle} {Proceedings of
  IEEE International Conference on Computers, Systems, and Signal
  Processing}}}\ (\bibinfo {address} {India},\ \bibinfo {year} {1984})\ p.\
  \bibinfo {pages} {175}\BibitemShut {NoStop}%
\bibitem [{\citenamefont {Nauerth}\ \emph {et~al.}(2013)\citenamefont
  {Nauerth}, \citenamefont {Moll}, \citenamefont {Rau}, \citenamefont {Fuchs},
  \citenamefont {Horwath}, \citenamefont {Frick},\ and\ \citenamefont
  {Weinfurter}}]{airbb84}%
  \BibitemOpen
  \bibfield  {author} {\bibinfo {author} {\bibfnamefont {S.}~\bibnamefont
  {Nauerth}}, \bibinfo {author} {\bibfnamefont {F.}~\bibnamefont {Moll}},
  \bibinfo {author} {\bibfnamefont {M.}~\bibnamefont {Rau}}, \bibinfo {author}
  {\bibfnamefont {C.}~\bibnamefont {Fuchs}}, \bibinfo {author} {\bibfnamefont
  {J.}~\bibnamefont {Horwath}}, \bibinfo {author} {\bibfnamefont
  {S.}~\bibnamefont {Frick}}, \ and\ \bibinfo {author} {\bibfnamefont
  {H.}~\bibnamefont {Weinfurter}},\ }\href@noop {} {\bibfield  {journal}
  {\bibinfo  {journal} {Nature Photonics}\ }\textbf {\bibinfo {volume} {7}},\
  \bibinfo {pages} {382} (\bibinfo {year} {2013})}\BibitemShut {NoStop}%
\bibitem [{\citenamefont {Ralph}(2000)}]{cv2}%
  \BibitemOpen
  \bibfield  {author} {\bibinfo {author} {\bibfnamefont {T.~C.}\ \bibnamefont
  {Ralph}},\ }\href {\doibase 10.1103/PhysRevA.62.062306} {\bibfield  {journal}
  {\bibinfo  {journal} {Phys. Rev. A}\ }\textbf {\bibinfo {volume} {62}},\
  \bibinfo {pages} {062306} (\bibinfo {year} {2000})}\BibitemShut {NoStop}%
\bibitem [{\citenamefont {Grosshans}\ and\ \citenamefont
  {Grangier}(2002)}]{cv3}%
  \BibitemOpen
  \bibfield  {author} {\bibinfo {author} {\bibfnamefont {F.}~\bibnamefont
  {Grosshans}}\ and\ \bibinfo {author} {\bibfnamefont {P.}~\bibnamefont
  {Grangier}},\ }\href {\doibase 10.1103/PhysRevLett.88.057902} {\bibfield
  {journal} {\bibinfo  {journal} {Phys. Rev. Lett.}\ }\textbf {\bibinfo
  {volume} {88}},\ \bibinfo {pages} {057902} (\bibinfo {year}
  {2002})}\BibitemShut {NoStop}%
\bibitem [{\citenamefont {Weedbrook}\ \emph {et~al.}(2012)\citenamefont
  {Weedbrook}, \citenamefont {Pirandola}, \citenamefont
  {Garc{\'\i}a-Patr{\'o}n}, \citenamefont {Cerf}, \citenamefont {Ralph},
  \citenamefont {Shapiro},\ and\ \citenamefont {Lloyd}}]{cv1}%
  \BibitemOpen
  \bibfield  {author} {\bibinfo {author} {\bibfnamefont {C.}~\bibnamefont
  {Weedbrook}}, \bibinfo {author} {\bibfnamefont {S.}~\bibnamefont
  {Pirandola}}, \bibinfo {author} {\bibfnamefont {R.}~\bibnamefont
  {Garc{\'\i}a-Patr{\'o}n}}, \bibinfo {author} {\bibfnamefont {N.~J.}\
  \bibnamefont {Cerf}}, \bibinfo {author} {\bibfnamefont {T.~C.}\ \bibnamefont
  {Ralph}}, \bibinfo {author} {\bibfnamefont {J.~H.}\ \bibnamefont {Shapiro}},
  \ and\ \bibinfo {author} {\bibfnamefont {S.}~\bibnamefont {Lloyd}},\
  }\href@noop {} {\bibfield  {journal} {\bibinfo  {journal} {Reviews of Modern
  Physics}\ }\textbf {\bibinfo {volume} {84}},\ \bibinfo {pages} {621}
  (\bibinfo {year} {2012})}\BibitemShut {NoStop}%
\bibitem [{\citenamefont {Grosshans}(2005)}]{atk1}%
  \BibitemOpen
  \bibfield  {author} {\bibinfo {author} {\bibfnamefont {F.}~\bibnamefont
  {Grosshans}},\ }\href@noop {} {\bibfield  {journal} {\bibinfo  {journal}
  {Physical review letters}\ }\textbf {\bibinfo {volume} {94}},\ \bibinfo
  {pages} {020504} (\bibinfo {year} {2005})}\BibitemShut {NoStop}%
\bibitem [{\citenamefont {Renner}\ and\ \citenamefont {Cirac}(2009)}]{atk2}%
  \BibitemOpen
  \bibfield  {author} {\bibinfo {author} {\bibfnamefont {R.}~\bibnamefont
  {Renner}}\ and\ \bibinfo {author} {\bibfnamefont {J.~I.}\ \bibnamefont
  {Cirac}},\ }\href {\doibase 10.1103/PhysRevLett.102.110504} {\bibfield
  {journal} {\bibinfo  {journal} {Phys. Rev. Lett.}\ }\textbf {\bibinfo
  {volume} {102}},\ \bibinfo {pages} {110504} (\bibinfo {year}
  {2009})}\BibitemShut {NoStop}%
\bibitem [{\citenamefont {Xu}\ \emph {et~al.}(2015)\citenamefont {Xu},
  \citenamefont {Wei}, \citenamefont {Sajeed}, \citenamefont {Kaiser},
  \citenamefont {Sun}, \citenamefont {Tang}, \citenamefont {Qian},
  \citenamefont {Makarov},\ and\ \citenamefont {Lo}}]{tap1}%
  \BibitemOpen
  \bibfield  {author} {\bibinfo {author} {\bibfnamefont {F.}~\bibnamefont
  {Xu}}, \bibinfo {author} {\bibfnamefont {K.}~\bibnamefont {Wei}}, \bibinfo
  {author} {\bibfnamefont {S.}~\bibnamefont {Sajeed}}, \bibinfo {author}
  {\bibfnamefont {S.}~\bibnamefont {Kaiser}}, \bibinfo {author} {\bibfnamefont
  {S.}~\bibnamefont {Sun}}, \bibinfo {author} {\bibfnamefont {Z.}~\bibnamefont
  {Tang}}, \bibinfo {author} {\bibfnamefont {L.}~\bibnamefont {Qian}}, \bibinfo
  {author} {\bibfnamefont {V.}~\bibnamefont {Makarov}}, \ and\ \bibinfo
  {author} {\bibfnamefont {H.-K.}\ \bibnamefont {Lo}},\ }\href@noop {}
  {\bibfield  {journal} {\bibinfo  {journal} {Physical Review A}\ }\textbf
  {\bibinfo {volume} {92}},\ \bibinfo {pages} {032305} (\bibinfo {year}
  {2015})}\BibitemShut {NoStop}%
\bibitem [{\citenamefont {Li}\ \emph {et~al.}(2011)\citenamefont {Li},
  \citenamefont {Wang}, \citenamefont {Huang}, \citenamefont {Chen},
  \citenamefont {Yin}, \citenamefont {Li}, \citenamefont {Zhou}, \citenamefont
  {Liu}, \citenamefont {Zhang}, \citenamefont {Guo} \emph {et~al.}}]{tap2}%
  \BibitemOpen
  \bibfield  {author} {\bibinfo {author} {\bibfnamefont {H.-W.}\ \bibnamefont
  {Li}}, \bibinfo {author} {\bibfnamefont {S.}~\bibnamefont {Wang}}, \bibinfo
  {author} {\bibfnamefont {J.-Z.}\ \bibnamefont {Huang}}, \bibinfo {author}
  {\bibfnamefont {W.}~\bibnamefont {Chen}}, \bibinfo {author} {\bibfnamefont
  {Z.-Q.}\ \bibnamefont {Yin}}, \bibinfo {author} {\bibfnamefont {F.-Y.}\
  \bibnamefont {Li}}, \bibinfo {author} {\bibfnamefont {Z.}~\bibnamefont
  {Zhou}}, \bibinfo {author} {\bibfnamefont {D.}~\bibnamefont {Liu}}, \bibinfo
  {author} {\bibfnamefont {Y.}~\bibnamefont {Zhang}}, \bibinfo {author}
  {\bibfnamefont {G.-C.}\ \bibnamefont {Guo}},  \emph {et~al.},\ }\href@noop {}
  {\bibfield  {journal} {\bibinfo  {journal} {Physical Review A}\ }\textbf
  {\bibinfo {volume} {84}},\ \bibinfo {pages} {062308} (\bibinfo {year}
  {2011})}\BibitemShut {NoStop}%
\bibitem [{\citenamefont {Zhao}\ \emph {et~al.}(2008)\citenamefont {Zhao},
  \citenamefont {Fung}, \citenamefont {Qi}, \citenamefont {Chen},\ and\
  \citenamefont {Lo}}]{tap3}%
  \BibitemOpen
  \bibfield  {author} {\bibinfo {author} {\bibfnamefont {Y.}~\bibnamefont
  {Zhao}}, \bibinfo {author} {\bibfnamefont {C.-H.~F.}\ \bibnamefont {Fung}},
  \bibinfo {author} {\bibfnamefont {B.}~\bibnamefont {Qi}}, \bibinfo {author}
  {\bibfnamefont {C.}~\bibnamefont {Chen}}, \ and\ \bibinfo {author}
  {\bibfnamefont {H.-K.}\ \bibnamefont {Lo}},\ }\href@noop {} {\bibfield
  {journal} {\bibinfo  {journal} {Physical Review A}\ }\textbf {\bibinfo
  {volume} {78}},\ \bibinfo {pages} {042333} (\bibinfo {year}
  {2008})}\BibitemShut {NoStop}%
\bibitem [{\citenamefont {Makarov}\ \emph {et~al.}(2006)\citenamefont
  {Makarov}, \citenamefont {Anisimov},\ and\ \citenamefont {Skaar}}]{tap4}%
  \BibitemOpen
  \bibfield  {author} {\bibinfo {author} {\bibfnamefont {V.}~\bibnamefont
  {Makarov}}, \bibinfo {author} {\bibfnamefont {A.}~\bibnamefont {Anisimov}}, \
  and\ \bibinfo {author} {\bibfnamefont {J.}~\bibnamefont {Skaar}},\
  }\href@noop {} {\bibfield  {journal} {\bibinfo  {journal} {Physical Review
  A}\ }\textbf {\bibinfo {volume} {74}},\ \bibinfo {pages} {022313} (\bibinfo
  {year} {2006})}\BibitemShut {NoStop}%
\bibitem [{\citenamefont {Lamas-Linares}\ and\ \citenamefont
  {Kurtsiefer}(2007)}]{tap5}%
  \BibitemOpen
  \bibfield  {author} {\bibinfo {author} {\bibfnamefont {A.}~\bibnamefont
  {Lamas-Linares}}\ and\ \bibinfo {author} {\bibfnamefont {C.}~\bibnamefont
  {Kurtsiefer}},\ }\href@noop {} {\bibfield  {journal} {\bibinfo  {journal}
  {Optics express}\ }\textbf {\bibinfo {volume} {15}},\ \bibinfo {pages} {9388}
  (\bibinfo {year} {2007})}\BibitemShut {NoStop}%
\bibitem [{\citenamefont {Lo}\ \emph {et~al.}(2012)\citenamefont {Lo},
  \citenamefont {Curty},\ and\ \citenamefont {Qi}}]{dvmdi}%
  \BibitemOpen
  \bibfield  {author} {\bibinfo {author} {\bibfnamefont {H.-K.}\ \bibnamefont
  {Lo}}, \bibinfo {author} {\bibfnamefont {M.}~\bibnamefont {Curty}}, \ and\
  \bibinfo {author} {\bibfnamefont {B.}~\bibnamefont {Qi}},\ }\href@noop {}
  {\bibfield  {journal} {\bibinfo  {journal} {Physical review letters}\
  }\textbf {\bibinfo {volume} {108}},\ \bibinfo {pages} {130503} (\bibinfo
  {year} {2012})}\BibitemShut {NoStop}%
\bibitem [{\citenamefont {Ac{\'\i}n}\ \emph {et~al.}(2007)\citenamefont
  {Ac{\'\i}n}, \citenamefont {Brunner}, \citenamefont {Gisin}, \citenamefont
  {Massar}, \citenamefont {Pironio},\ and\ \citenamefont {Scarani}}]{dvdi}%
  \BibitemOpen
  \bibfield  {author} {\bibinfo {author} {\bibfnamefont {A.}~\bibnamefont
  {Ac{\'\i}n}}, \bibinfo {author} {\bibfnamefont {N.}~\bibnamefont {Brunner}},
  \bibinfo {author} {\bibfnamefont {N.}~\bibnamefont {Gisin}}, \bibinfo
  {author} {\bibfnamefont {S.}~\bibnamefont {Massar}}, \bibinfo {author}
  {\bibfnamefont {S.}~\bibnamefont {Pironio}}, \ and\ \bibinfo {author}
  {\bibfnamefont {V.}~\bibnamefont {Scarani}},\ }\href@noop {} {\bibfield
  {journal} {\bibinfo  {journal} {Physical Review Letters}\ }\textbf {\bibinfo
  {volume} {98}},\ \bibinfo {pages} {230501} (\bibinfo {year}
  {2007})}\BibitemShut {NoStop}%
\bibitem [{\citenamefont {Yang}\ and\ \citenamefont {Li}(2009)}]{pas}%
  \BibitemOpen
  \bibfield  {author} {\bibinfo {author} {\bibfnamefont {Y.}~\bibnamefont
  {Yang}}\ and\ \bibinfo {author} {\bibfnamefont {F.-L.}\ \bibnamefont {Li}},\
  }\href {\doibase 10.1103/PhysRevA.80.022315} {\bibfield  {journal} {\bibinfo
  {journal} {Phys. Rev. A}\ }\textbf {\bibinfo {volume} {80}},\ \bibinfo
  {pages} {022315} (\bibinfo {year} {2009})}\BibitemShut {NoStop}%
\bibitem [{\citenamefont {Lee}\ \emph {et~al.}(2011)\citenamefont {Lee},
  \citenamefont {Ji}, \citenamefont {Kim},\ and\ \citenamefont {Nha}}]{rep1}%
  \BibitemOpen
  \bibfield  {author} {\bibinfo {author} {\bibfnamefont {S.-Y.}\ \bibnamefont
  {Lee}}, \bibinfo {author} {\bibfnamefont {S.-W.}\ \bibnamefont {Ji}},
  \bibinfo {author} {\bibfnamefont {H.-J.}\ \bibnamefont {Kim}}, \ and\
  \bibinfo {author} {\bibfnamefont {H.}~\bibnamefont {Nha}},\ }\href {\doibase
  10.1103/PhysRevA.84.012302} {\bibfield  {journal} {\bibinfo  {journal} {Phys.
  Rev. A}\ }\textbf {\bibinfo {volume} {84}},\ \bibinfo {pages} {012302}
  (\bibinfo {year} {2011})}\BibitemShut {NoStop}%
\bibitem [{\citenamefont {Ourjoumtsev}\ \emph {et~al.}(2007)\citenamefont
  {Ourjoumtsev}, \citenamefont {Dantan}, \citenamefont {Tualle-Brouri},\ and\
  \citenamefont {Grangier}}]{ngexp1}%
  \BibitemOpen
  \bibfield  {author} {\bibinfo {author} {\bibfnamefont {A.}~\bibnamefont
  {Ourjoumtsev}}, \bibinfo {author} {\bibfnamefont {A.}~\bibnamefont {Dantan}},
  \bibinfo {author} {\bibfnamefont {R.}~\bibnamefont {Tualle-Brouri}}, \ and\
  \bibinfo {author} {\bibfnamefont {P.}~\bibnamefont {Grangier}},\ }\href
  {\doibase 10.1103/PhysRevLett.98.030502} {\bibfield  {journal} {\bibinfo
  {journal} {Phys. Rev. Lett.}\ }\textbf {\bibinfo {volume} {98}},\ \bibinfo
  {pages} {030502} (\bibinfo {year} {2007})}\BibitemShut {NoStop}%
\bibitem [{\citenamefont {Takahashi}\ \emph {et~al.}(2010)\citenamefont
  {Takahashi}, \citenamefont {Neergaard-Nielsen}, \citenamefont {Takeuchi},
  \citenamefont {Takeoka}, \citenamefont {Hayasaka}, \citenamefont {Furusawa},\
  and\ \citenamefont {Sasaki}}]{ngexp2}%
  \BibitemOpen
  \bibfield  {author} {\bibinfo {author} {\bibfnamefont {H.}~\bibnamefont
  {Takahashi}}, \bibinfo {author} {\bibfnamefont {J.~S.}\ \bibnamefont
  {Neergaard-Nielsen}}, \bibinfo {author} {\bibfnamefont {M.}~\bibnamefont
  {Takeuchi}}, \bibinfo {author} {\bibfnamefont {M.}~\bibnamefont {Takeoka}},
  \bibinfo {author} {\bibfnamefont {K.}~\bibnamefont {Hayasaka}}, \bibinfo
  {author} {\bibfnamefont {A.}~\bibnamefont {Furusawa}}, \ and\ \bibinfo
  {author} {\bibfnamefont {M.}~\bibnamefont {Sasaki}},\ }\href@noop {}
  {\bibfield  {journal} {\bibinfo  {journal} {Nature photonics}\ }\textbf
  {\bibinfo {volume} {4}},\ \bibinfo {pages} {178} (\bibinfo {year}
  {2010})}\BibitemShut {NoStop}%
\bibitem [{\citenamefont {Kurochkin}\ \emph {et~al.}(2014)\citenamefont
  {Kurochkin}, \citenamefont {Prasad},\ and\ \citenamefont {Lvovsky}}]{ngexp3}%
  \BibitemOpen
  \bibfield  {author} {\bibinfo {author} {\bibfnamefont {Y.}~\bibnamefont
  {Kurochkin}}, \bibinfo {author} {\bibfnamefont {A.~S.}\ \bibnamefont
  {Prasad}}, \ and\ \bibinfo {author} {\bibfnamefont {A.~I.}\ \bibnamefont
  {Lvovsky}},\ }\href {\doibase 10.1103/PhysRevLett.112.070402} {\bibfield
  {journal} {\bibinfo  {journal} {Phys. Rev. Lett.}\ }\textbf {\bibinfo
  {volume} {112}},\ \bibinfo {pages} {070402} (\bibinfo {year}
  {2014})}\BibitemShut {NoStop}%
\bibitem [{\citenamefont {Boyd}\ \emph {et~al.}(2007)\citenamefont {Boyd},
  \citenamefont {Chan},\ and\ \citenamefont {O'Sullivan}}]{pasexp}%
  \BibitemOpen
  \bibfield  {author} {\bibinfo {author} {\bibfnamefont {R.~W.}\ \bibnamefont
  {Boyd}}, \bibinfo {author} {\bibfnamefont {K.~W.~C.}\ \bibnamefont {Chan}}, \
  and\ \bibinfo {author} {\bibfnamefont {M.~N.}\ \bibnamefont {O'Sullivan}},\
  }\href {\doibase 10.1126/science.1148947} {\bibfield  {journal} {\bibinfo
  {journal} {Science}\ }\textbf {\bibinfo {volume} {317}},\ \bibinfo {pages}
  {1874} (\bibinfo {year} {2007})},\ \Eprint
  {http://arxiv.org/abs/https://www.science.org/doi/pdf/10.1126/science.1148947}
  {https://www.science.org/doi/pdf/10.1126/science.1148947} \BibitemShut
  {NoStop}%
\bibitem [{\citenamefont {Bartley}\ and\ \citenamefont
  {Walmsley}(2015)}]{bartley}%
  \BibitemOpen
  \bibfield  {author} {\bibinfo {author} {\bibfnamefont {T.~J.}\ \bibnamefont
  {Bartley}}\ and\ \bibinfo {author} {\bibfnamefont {I.~A.}\ \bibnamefont
  {Walmsley}},\ }\href@noop {} {\bibfield  {journal} {\bibinfo  {journal} {New
  Journal of Physics}\ }\textbf {\bibinfo {volume} {17}},\ \bibinfo {pages}
  {023038} (\bibinfo {year} {2015})}\BibitemShut {NoStop}%
\bibitem [{\citenamefont {Hu}\ \emph {et~al.}(2020)\citenamefont {Hu},
  \citenamefont {Al-amri}, \citenamefont {Liao},\ and\ \citenamefont
  {Zubairy}}]{zubairi1}%
  \BibitemOpen
  \bibfield  {author} {\bibinfo {author} {\bibfnamefont {L.}~\bibnamefont
  {Hu}}, \bibinfo {author} {\bibfnamefont {M.}~\bibnamefont {Al-amri}},
  \bibinfo {author} {\bibfnamefont {Z.}~\bibnamefont {Liao}}, \ and\ \bibinfo
  {author} {\bibfnamefont {M.~S.}\ \bibnamefont {Zubairy}},\ }\href {\doibase
  10.1103/PhysRevA.102.012608} {\bibfield  {journal} {\bibinfo  {journal}
  {Phys. Rev. A}\ }\textbf {\bibinfo {volume} {102}},\ \bibinfo {pages}
  {012608} (\bibinfo {year} {2020})}\BibitemShut {NoStop}%
\bibitem [{\citenamefont {Ma}\ \emph {et~al.}(2018)\citenamefont {Ma},
  \citenamefont {Huang}, \citenamefont {Bai}, \citenamefont {Wang},
  \citenamefont {Bao},\ and\ \citenamefont {Zeng}}]{mdisps}%
  \BibitemOpen
  \bibfield  {author} {\bibinfo {author} {\bibfnamefont {H.-X.}\ \bibnamefont
  {Ma}}, \bibinfo {author} {\bibfnamefont {P.}~\bibnamefont {Huang}}, \bibinfo
  {author} {\bibfnamefont {D.-Y.}\ \bibnamefont {Bai}}, \bibinfo {author}
  {\bibfnamefont {S.-Y.}\ \bibnamefont {Wang}}, \bibinfo {author}
  {\bibfnamefont {W.-S.}\ \bibnamefont {Bao}}, \ and\ \bibinfo {author}
  {\bibfnamefont {G.-H.}\ \bibnamefont {Zeng}},\ }\href@noop {} {\bibfield
  {journal} {\bibinfo  {journal} {Physical Review A}\ }\textbf {\bibinfo
  {volume} {97}},\ \bibinfo {pages} {042329} (\bibinfo {year}
  {2018})}\BibitemShut {NoStop}%
\bibitem [{\citenamefont {Ye}\ \emph {et~al.}(2020)\citenamefont {Ye},
  \citenamefont {Zhong}, \citenamefont {Wu}, \citenamefont {Hu},\ and\
  \citenamefont {Guo}}]{mdizpc}%
  \BibitemOpen
  \bibfield  {author} {\bibinfo {author} {\bibfnamefont {W.}~\bibnamefont
  {Ye}}, \bibinfo {author} {\bibfnamefont {H.}~\bibnamefont {Zhong}}, \bibinfo
  {author} {\bibfnamefont {X.}~\bibnamefont {Wu}}, \bibinfo {author}
  {\bibfnamefont {L.}~\bibnamefont {Hu}}, \ and\ \bibinfo {author}
  {\bibfnamefont {Y.}~\bibnamefont {Guo}},\ }\href@noop {} {\bibfield
  {journal} {\bibinfo  {journal} {Quantum Information Processing}\ }\textbf
  {\bibinfo {volume} {19}},\ \bibinfo {pages} {1} (\bibinfo {year}
  {2020})}\BibitemShut {NoStop}%
\bibitem [{\citenamefont {Singh}\ and\ \citenamefont {Bose}(2021)}]{jsingh}%
  \BibitemOpen
  \bibfield  {author} {\bibinfo {author} {\bibfnamefont {J.}~\bibnamefont
  {Singh}}\ and\ \bibinfo {author} {\bibfnamefont {S.}~\bibnamefont {Bose}},\
  }\href {\doibase 10.1103/PhysRevA.104.052605} {\bibfield  {journal} {\bibinfo
   {journal} {Phys. Rev. A}\ }\textbf {\bibinfo {volume} {104}},\ \bibinfo
  {pages} {052605} (\bibinfo {year} {2021})}\BibitemShut {NoStop}%
\bibitem [{\citenamefont {Khan}\ \emph {et~al.}(2023)\citenamefont {Khan},
  \citenamefont {Waseem}, \citenamefont {Irfan}, \citenamefont {Mehmood},\ and\
  \citenamefont {Qamar}}]{zpc8mdi}%
  \BibitemOpen
  \bibfield  {author} {\bibinfo {author} {\bibfnamefont {M.~B.}\ \bibnamefont
  {Khan}}, \bibinfo {author} {\bibfnamefont {M.}~\bibnamefont {Waseem}},
  \bibinfo {author} {\bibfnamefont {M.}~\bibnamefont {Irfan}}, \bibinfo
  {author} {\bibfnamefont {A.}~\bibnamefont {Mehmood}}, \ and\ \bibinfo
  {author} {\bibfnamefont {S.}~\bibnamefont {Qamar}},\ }\href {\doibase
  10.1364/JOSAB.482577} {\bibfield  {journal} {\bibinfo  {journal} {J. Opt.
  Soc. Am. B}\ }\textbf {\bibinfo {volume} {40}},\ \bibinfo {pages} {763}
  (\bibinfo {year} {2023})}\BibitemShut {NoStop}%
\bibitem [{\citenamefont {Li}\ \emph {et~al.}(2014)\citenamefont {Li},
  \citenamefont {Zhang}, \citenamefont {Xu}, \citenamefont {Peng},\ and\
  \citenamefont {Guo}}]{cvmdi}%
  \BibitemOpen
  \bibfield  {author} {\bibinfo {author} {\bibfnamefont {Z.}~\bibnamefont
  {Li}}, \bibinfo {author} {\bibfnamefont {Y.-C.}\ \bibnamefont {Zhang}},
  \bibinfo {author} {\bibfnamefont {F.}~\bibnamefont {Xu}}, \bibinfo {author}
  {\bibfnamefont {X.}~\bibnamefont {Peng}}, \ and\ \bibinfo {author}
  {\bibfnamefont {H.}~\bibnamefont {Guo}},\ }\href@noop {} {\bibfield
  {journal} {\bibinfo  {journal} {Physical Review A}\ }\textbf {\bibinfo
  {volume} {89}},\ \bibinfo {pages} {052301} (\bibinfo {year}
  {2014})}\BibitemShut {NoStop}%
\bibitem [{\citenamefont {Vahlbruch}\ \emph {et~al.}(2016)\citenamefont
  {Vahlbruch}, \citenamefont {Mehmet}, \citenamefont {Danzmann},\ and\
  \citenamefont {Schnabel}}]{sqdb1}%
  \BibitemOpen
  \bibfield  {author} {\bibinfo {author} {\bibfnamefont {H.}~\bibnamefont
  {Vahlbruch}}, \bibinfo {author} {\bibfnamefont {M.}~\bibnamefont {Mehmet}},
  \bibinfo {author} {\bibfnamefont {K.}~\bibnamefont {Danzmann}}, \ and\
  \bibinfo {author} {\bibfnamefont {R.}~\bibnamefont {Schnabel}},\ }\href
  {\doibase 10.1103/PhysRevLett.117.110801} {\bibfield  {journal} {\bibinfo
  {journal} {Phys. Rev. Lett.}\ }\textbf {\bibinfo {volume} {117}},\ \bibinfo
  {pages} {110801} (\bibinfo {year} {2016})}\BibitemShut {NoStop}%
\bibitem [{\citenamefont {Ma}\ \emph {et~al.}(2019)\citenamefont {Ma},
  \citenamefont {Li}, \citenamefont {Xie},\ and\ \citenamefont {Li}}]{sqdb2}%
  \BibitemOpen
  \bibfield  {author} {\bibinfo {author} {\bibfnamefont {S.-l.}\ \bibnamefont
  {Ma}}, \bibinfo {author} {\bibfnamefont {X.-k.}\ \bibnamefont {Li}}, \bibinfo
  {author} {\bibfnamefont {J.-k.}\ \bibnamefont {Xie}}, \ and\ \bibinfo
  {author} {\bibfnamefont {F.-l.}\ \bibnamefont {Li}},\ }\href {\doibase
  10.1103/PhysRevA.99.012325} {\bibfield  {journal} {\bibinfo  {journal} {Phys.
  Rev. A}\ }\textbf {\bibinfo {volume} {99}},\ \bibinfo {pages} {012325}
  (\bibinfo {year} {2019})}\BibitemShut {NoStop}%
\bibitem [{\citenamefont {M\o{}lmer}(2006)}]{sqdb3}%
  \BibitemOpen
  \bibfield  {author} {\bibinfo {author} {\bibfnamefont {K.}~\bibnamefont
  {M\o{}lmer}},\ }\href {\doibase 10.1103/PhysRevA.73.063804} {\bibfield
  {journal} {\bibinfo  {journal} {Phys. Rev. A}\ }\textbf {\bibinfo {volume}
  {73}},\ \bibinfo {pages} {063804} (\bibinfo {year} {2006})}\BibitemShut
  {NoStop}%
\bibitem [{\citenamefont {Hosseinidehaj}(2017)}]{phdthesis}%
  \BibitemOpen
  \bibfield  {author} {\bibinfo {author} {\bibfnamefont {N.}~\bibnamefont
  {Hosseinidehaj}},\ }\emph {\bibinfo {title} {Continuous-Variable Quantum
  Communication over Free-Space Lossy Channels}},\ \href@noop {} {Ph.D.
  thesis},\ \bibinfo  {school} {The University of New South Wales} (\bibinfo
  {year} {2017})\BibitemShut {NoStop}%
\bibitem [{\citenamefont {Vidal}\ and\ \citenamefont {Werner}(2002)}]{logn1}%
  \BibitemOpen
  \bibfield  {author} {\bibinfo {author} {\bibfnamefont {G.}~\bibnamefont
  {Vidal}}\ and\ \bibinfo {author} {\bibfnamefont {R.~F.}\ \bibnamefont
  {Werner}},\ }\href {\doibase 10.1103/PhysRevA.65.032314} {\bibfield
  {journal} {\bibinfo  {journal} {Phys. Rev. A}\ }\textbf {\bibinfo {volume}
  {65}},\ \bibinfo {pages} {032314} (\bibinfo {year} {2002})}\BibitemShut
  {NoStop}%
\bibitem [{\citenamefont {Plenio}(2005)}]{logn2}%
  \BibitemOpen
  \bibfield  {author} {\bibinfo {author} {\bibfnamefont {M.~B.}\ \bibnamefont
  {Plenio}},\ }\href {\doibase 10.1103/PhysRevLett.95.090503} {\bibfield
  {journal} {\bibinfo  {journal} {Phys. Rev. Lett.}\ }\textbf {\bibinfo
  {volume} {95}},\ \bibinfo {pages} {090503} (\bibinfo {year}
  {2005})}\BibitemShut {NoStop}%
\bibitem [{\citenamefont {Villasenor}\ and\ \citenamefont
  {Malaney}(2019)}]{lowsq}%
  \BibitemOpen
  \bibfield  {author} {\bibinfo {author} {\bibfnamefont {E.}~\bibnamefont
  {Villasenor}}\ and\ \bibinfo {author} {\bibfnamefont {R.}~\bibnamefont
  {Malaney}},\ }in\ \href {\doibase 10.1109/GCWkshps45667.2019.9024548} {\emph
  {\bibinfo {booktitle} {2019 IEEE Globecom Workshops (GC Wkshps)}}}\ (\bibinfo
  {year} {2019})\ pp.\ \bibinfo {pages} {1--6}\BibitemShut {NoStop}%
\bibitem [{\citenamefont {Garc\'{\i}a-Patr\'on}\ and\ \citenamefont
  {Cerf}(2006)}]{optimalityofG}%
  \BibitemOpen
  \bibfield  {author} {\bibinfo {author} {\bibfnamefont {R.}~\bibnamefont
  {Garc\'{\i}a-Patr\'on}}\ and\ \bibinfo {author} {\bibfnamefont {N.~J.}\
  \bibnamefont {Cerf}},\ }\href {\doibase 10.1103/PhysRevLett.97.190503}
  {\bibfield  {journal} {\bibinfo  {journal} {Phys. Rev. Lett.}\ }\textbf
  {\bibinfo {volume} {97}},\ \bibinfo {pages} {190503} (\bibinfo {year}
  {2006})}\BibitemShut {NoStop}%
\end{thebibliography}%

\end{document}